\documentclass{PoS}

\usepackage{amsmath}

\title{Theta dependence in the large N limit}

\ShortTitle{Theta dependence in the large N limit}

\author{\speaker{Claudio Bonati}%
        \thanks{Present address: \newline
        Dipartimento di Fisica e Astronomia, Universit\`a di Firenze and INFN, Sezione di Firenze,\newline 
        Via Sansone 1, 50019 Sesto Fiorentino (FI)}\\
        Dipartimento di Fisica, Universit\`a di Pisa and INFN, Sezione di Pisa \\
        Largo Pontecorvo 3, 56127 Pisa, Italy
        E-mail: \email{claudio.bonati@df.unipi.it}}

\author{Massimo D'Elia \\
        Dipartimento di Fisica, Universit\`a di Pisa and INFN, Sezione di Pisa \\
        Largo Pontecorvo 3, 56127 Pisa, Italy
        E-mail: \email{massimo.delia@unipi.it}}

\author{Paolo Rossi \\
        Dipartimento di Fisica, Universit\`a di Pisa and INFN, Sezione di Pisa \\
        Largo Pontecorvo 3, 56127 Pisa, Italy
        E-mail: \email{paolo.rossi@unipi.it}}

\author{Ettore Vicari \\
        Dipartimento di Fisica, Universit\`a di Pisa and INFN, Sezione di Pisa \\
        Largo Pontecorvo 3, 56127 Pisa, Italy
        E-mail: \email{ettore.vicari@unipi.it}}

\abstract{
Studies of the large $N$ behaviour of the topological properties of gauge
theories typically focused on the large $N$ scaling of the topological
susceptibility. A much more difficult task is the study of the behaviour of
higher cumulants of the topological charge in the large $N$ limit, which up to
now remained elusive. We will present first results confirming the expected
large $N$ scaling of the coefficient commonly denoted by $b_2$, related to the
kurtosis of the topological charge.
}

\FullConference{34th annual International Symposium on Lattice Field Theory\\
         24-30 July 2016\\
         University of Southampton, UK}

\begin{document}

\section{Introduction}\label{sec:intro}

In recent times there has been a renewed interest in the $\theta$ dependence of
gauge theories: this topic emerges naturally in several approaches to the
physics of strongly interacting matter, both theoretically oriented (like
semiclassical methods, expansion in the number of colors, holographic and
lattice methods) and phenomenologically relevant (like $U_A(1)$ problem,
$\eta'$ physics and axions).  The euclidean Lagrangian of gauge theories in
the presence of a non-vanishing $\theta$ parameter is
\begin{equation}\label{lagrangian}
\mathcal{L}_\theta  = \frac{1}{4} F_{\mu\nu}^a(x)F_{\mu\nu}^a(x)
- i \theta q(x)\ ,\qquad 
q(x)=\frac{g^2}{64\pi^2} 
\epsilon_{\mu\nu\rho\sigma} F_{\mu\nu}^a(x) F_{\rho\sigma}^a(x)\ ,
\end{equation}
where $q(x)$ is the topological charge density, whose four-dimensional integral
is (for smooth configurations with finite action) an integer number: the
topological charge $Q$.

Since $q(x)$ can be written as the four-divergence of the Chern-Simons current,
the theory is in fact independent of $\theta$ both at the classical and at the
perturbative quantum level; nevertheless nonperturbative quantum effects induce
a dependence of observables on the $\theta$ value. Of particular interest is
the dependence on $\theta$ of the ground state energy density $E$ (or at finite
temperature of the free energy density $F$), which can be parametrized in the form
(see e.g. \cite{Vicari:2008jw})
\begin{equation}\label{thdep}
E(\theta)-E(0)=\frac{1}{2}\chi\theta^2\big(1+b_2\theta^2+b_4\theta^4+\cdots\big)\ ,
\end{equation}
where $\chi$ is, by definition, the topological susceptibility of the theory
and the coefficients $b_{2n}$s characterize deviations from the leading
quadratic behaviour.  

This expansion is expected to have a finite radius of convergence, however no
generic analytical method is known to compute the coefficients $\chi, b_{2n}$s
from first principles with a systematically improvable
precision\footnote{Methods exist to deal with two specific cases: the case in
which light quarks and spontaneous chiral symmetry breaking are present
\cite{DiVecchia:1980ve} and the case of asymptotically hight temperature
\cite{Gross:1980br}.}. A systematic framework for evaluating $\chi$ and
$b_{2n}$s by means of numerical simulations is instead provided by the lattice
discretization of the theory, although some difficulties are encountered, as
will be discussed in the next sections.

A drastic simplification of the $\theta$ dependence of the ground state energy
density happens in the limit of an infinite number of colors: it is in fact
quite easy to obtain, using standard large $N$ scaling arguments (see
\cite{Vicari:2008jw, WittenLargeN}), the relations
\begin{eqnarray}\label{largeNsun}
\chi(N)=\bar{\chi}+O(N^{-2}),\quad b_{2n}(N)=\bar{b}_{2n}N^{-2n}+O(N^{-2n-2})\ , 
\end{eqnarray}
where $\bar{\chi}$ and $\bar{b}_{2n}$ are $N$-independent numbers. From these
relations it follows that, in the limit of a large number of colors,
$\chi\to\bar{\chi}$ and $b_{2n}\to 0$. It has however to be explicitly remarked
that the relations Eq~\eqref{largeNsun} are not expected to be universally
valid: they are obtained by assuming that $\bar{\theta}=\theta/N$ is the correct
scaling variable in the large $N$ limit, which follows from the assumption
that $\lim_{N\to\infty}\chi(N)$ is not singular (i.e. not vanishing nor
divergent).  When numerically verifying the first and the second of
Eq.~\eqref{largeNsun} one is in fact checking two different aspects: by
checking that the first equation is satisfied with $0<\bar{\chi}<\infty$ one
confirms the basic hypothesis used to study the $\theta$ dependence in the
large $N$ limit (and that is needed for the solution of the $U_A(1)$ problem),
when checking the second relation one is verifying the internal consistency of
this hypothesis.

The main aim of all the past studies concerning the large $N$ behaviour of the
$\theta$ dependence was the study of $\chi(N)$; a notable exception is the
study by Del Debbio et al. in \cite{chiLargeN}, in which first results for
$b_2$ in the $N=3,4, 6$ cases were reported, but the precision was too poor to
draw firm conclusions from data. On the other hand several studies later
investigated the value of $b_2$ for the case of $SU(3)$, obtaining nicely
compatible results and reaching a relative precision around $7\%$ (see Fig.~6
of \cite{Bonati:2015sqt}). In the following we will discuss the main ideas and
results of the work \cite{Bonati:2016tvi} (to which we refer for more details),
in which the scaling relation for $b_2$ in Eq~\eqref{largeNsun} was for the
first time numerically confirmed.

\section{Numerical setup}\label{sec:numset}

The standard way of computing the coefficients $\chi$ and $b_{2n}$s appearing
in Eq.~\eqref{thdep} is to study the cumulants of the distribution of the
topological charge $Q$ at $\theta=0$. It is indeed easy to show that the lowest
order coefficients are given by (odd momenta vanish because of the $CP$
invariance at $\theta=0$)
\begin{equation}\label{theta0method}
\begin{aligned}
& \chi = \frac{\langle Q^2 \rangle_{\theta=0}}{\mathcal{V}},\quad 
b_2=-\frac{\langle Q^4\rangle_{\theta=0}-
         3\langle Q^2\rangle^2_{\theta=0}}{12\langle Q^2\rangle_{\theta=0}}, \\
& b_4=\frac{\left[ \langle Q^6\rangle-15\langle Q^2\rangle \langle Q^4\rangle 
+30\langle Q^2\rangle ^3\right]_{\theta=0}}{360 \langle Q^2\rangle_{\theta=0}} \ ,
\end{aligned}
\end{equation}
where $\mathcal{V}$ is the four-dimensional volume.  From these expressions it
follows that the topological susceptibility is the variance of the $Q$
distribution, while the $b_{2n}$s coefficients parametrize the deviations from
a Gaussian distribution. 

The last sentence often causes some confusion due to the fact that, by the
central limit theorem, the distribution of $Q$ becomes closer and closer to a
Gaussian in the thermodynamic limit and one could erroneously guess the
$b_{2n}$s coefficients to vanish in this limit.  However the central limit
theorem just states that the probability distribution of the variable
$Q/\sqrt{\mathcal{V}}$ pointwise converges to a Gaussian distribution, and this
does not imply that the cumulants of the non rescaled variable $Q$ (the ones
appearing in the numerators of Eq.~\eqref{theta0method}) vanishes. In fact
these cumulants are extensive, in such a way that the $b_{2n}$s are intensive
quantities, as should be clear from Eq.~\eqref{thdep}.

%However there is no
%contradiction between the central limit theorem and a non-vanishing value of the
%$b_{2n}$s coefficients: the central limit theorem states that the probability distributions
%of the variable $Q/\sqrt{\mathcal{V}}$ converges pointwise to a Gaussian,  

%it is just a matter of performing the operations in the
%correct order: to proceed correctly we have \emph{first} to compute the
%averages appearing in Eq.~\eqref{theta0method} at fixed volume and \emph{then}
%to perform the thermodynamic limit of the result. 

The central limit theorem however has important consequences on the scaling
with the volume of Monte Carlo errors: it is intuitively clear that the estimator of a
quantity measuring the deviation from a Gaussian distribution will be noisier
and noisier as the volume is increased. This can be easily formalized (see e.g.
\cite{Bonati:2015sqt}) and the outcome is that, at fixed Monte Carlo
statistics, the error of the $b_{2n}$ estimator at $\theta=0$ grows with the volume like
$\mathcal{V}^n$.  The way out of this lacking of self-averaging is well know
\cite{MBH}: instead of evaluating fluctuation observables at vanishing external
field, one has to study the response to an external source. In the present
context, an external source is a non-zero value of the $\theta$ parameter, that
has to be imaginary in order not to spoil the reality of the action:
$\theta=-i\theta_I$. It is then easy to show that \cite{Panagopoulos:2011rb} 
\begin{equation}\label{Qcum1th}
\frac{\langle Q\rangle_{\theta_I}}{\mathcal{V}}=\chi\theta_I(1-2b_2\theta_I^2+
3b_4\theta_I^4+\ldots )\ ,
\end{equation}
from which it follows that $\chi$ and the $b_{2n}$s coefficients can be
extracted, e.g.,  from the $\theta$ dependence of the average $\langle
Q\rangle_{\theta_I}$, whose error does not grow with the volume.

\section{Numerical results}\label{sec:numres}

The discretized action adopted in the simulations was
\begin{equation}\label{latticeaction}
S[U] = S_W[U] - \theta_{L} Q_L[U] \, ,
\end{equation}
where $S_W$ is the Wilson action, $Q_L=\sum_x q_L(x)$ and $q_L(x)$ is
the simplest discretization of the topological charge density with definite
parity:
\begin{equation}\label{qlattice}
q_L(x) = -\frac{1}{2^9 \pi^2} 
\sum_{\mu\nu\rho\sigma = \pm 1}^{\pm 4} 
{\tilde{\epsilon}}_{\mu\nu\rho\sigma} \hbox{Tr} \left( 
\Pi_{\mu\nu}(x) \Pi_{\rho\sigma}(x) \right) \ . 
\end{equation}
In the last expression $\Pi_{\mu\nu}$ denotes the plaquette while
$\tilde{\epsilon}_{\mu\nu\rho\sigma}$ is an extension of the usual Levi-Civita
tensor, defined for negative entries by ${\tilde{\epsilon}}_{\mu\nu\rho\sigma}
= -{\tilde{\epsilon}}_{(-\mu)\nu\rho\sigma}$ and complete antisymmetry. This
simple discretization makes the MC update easier, however it has the
disadvantage of inducing a finite renormalization of the lattice operator $q_L$
\cite{Campostrini:1988cy}, which translate in a finite renormalization of the lattice
$\theta$ parameter: $\theta_I=Z(a)\theta_L$ (where $a$ is the lattice spacing).

To avoid the appearance of further renormalizations also in the observables,
the topological charge $Q_L$ was measured after cooling \cite{cooling} (see
\cite{smoothers} for discussions about the equivalence of different smoothing
algorithms, like gradient-flow), and in particular the following prescription
was used to assign to a configuration a value of the topological charge:
\begin{equation}\label{Q_round}
Q=\mathrm{round}\left(\alpha\, Q_L^{\rm cool} \right)\ , 
\end{equation} 
where $\mathrm{round}(\cdot)$ denotes the truncation to the closest integer and
the coefficient $\alpha$ is defined in such a way to make $\alpha Q_L^{\rm
cool}$ on average as close as possible to the integer values (see
\cite{Bonati:2016tvi} for more details).

Measures have been performed after several cooling steps (ranging from 5 to
25), in order to check for the stability of the results, and the values of $Z$,
$\chi$, $b_2$ and $b_4$ have been extracted by fitting the first four cumulants
of $Q$ according to the relations:
\begin{equation}\label{eq:cum_theta_L}
\begin{split}
\frac{\langle Q \rangle_{\theta_L}}{\mathcal{V}}&=
\chi Z \theta_L (1 - 2 b_2 Z^2 \theta_L^2 + 3 b_4 Z^4 \theta_L^4 + \dots)\, , \\
\frac{\langle Q^2 \rangle_{\theta_L, c}}{\mathcal{V}}&=  
\chi (1 - 6 b_2 Z^2 \theta_L^2 + 15 b_4 Z^4 \theta_L^4 + \dots)\, , \\
\frac{\langle Q^3 \rangle_{\theta_L, c}}{\mathcal{V}}&=  
\chi (- 12 b_2 Z \theta_L + 60 b_4 Z^3 \theta_L^3 + \dots)\, , \\
\frac{\langle Q^4 \rangle_{\theta_L, c}}{\mathcal{V}}&=
\chi (- 12 b_2 + 180 b_4 Z^2 \theta_L^2 + \dots)\, .
\end{split}
\end{equation}

\begin{figure}[t]
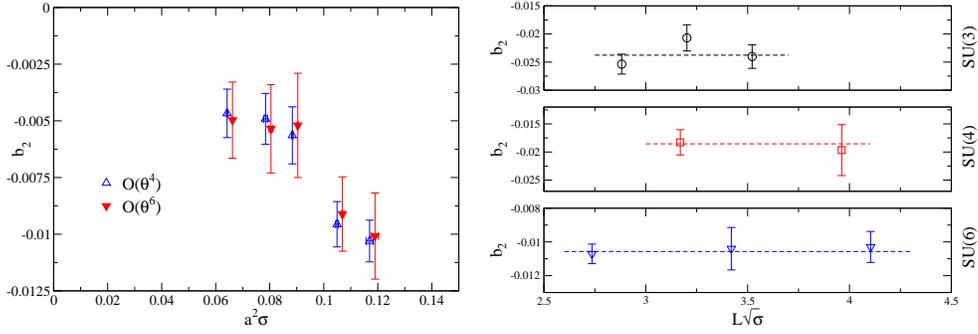

\centering
\includegraphics[width=0.40\textwidth,clip]{./b2_su6.eps}\quad  
\includegraphics[width=0.43\textwidth,clip]{./b2_thlimit.eps}
\caption{Tests for systematics: (left) dependence of the $SU(6)$ $b_2$ results
on the truncation adopted in the fitting procedure; (right) dependence of $b_2$
on the lattice size adopted (at fixed lattice spacing).}
\label{fig:sistematici}
\end{figure}

In order to check for systematics, different truncations of these equations
have been tested. By keeping all the terms up to $\theta^6$ in the expansion of
$E(\theta)$ it was possible to obtain estimates for all the coefficients up to
$b_4$ (with the $b_4$ values that always turned out to be compatible with
zero).  Compatible results for $Z$, $\chi$ and $b_2$ were obtained in all the
cases by using a truncation to $O(\theta^4)$, see Fig.~\ref{fig:sistematici}
(left) for the example of the $SU(6)$ case. The use of the imaginary $\theta$
source enabled us to reach very large physical volumes (up to
$L\sqrt{\sigma}\gtrsim 4$) and to exclude the presence of any sizable finite
volume effects, see Fig.~\ref{fig:sistematici} (right). 

To continuum extrapolate the results we used a fit adopting the leading
$O(a^2)$ correction and checking for systematics by varying the fit range.
This was done in all the cases but for $b_2$ in the $SU(6)$ case; in this case
no sizable dependence on lattice spacing was observed for $a^2\sigma\lesssim
0.1$ (that was the continuum scaling region used for the other $N$ values) and
the conservative estimate $-0.0045(15)$ was used (for more details and tables of 
numerical values see \cite{Bonati:2016tvi}).

\begin{figure}[b]
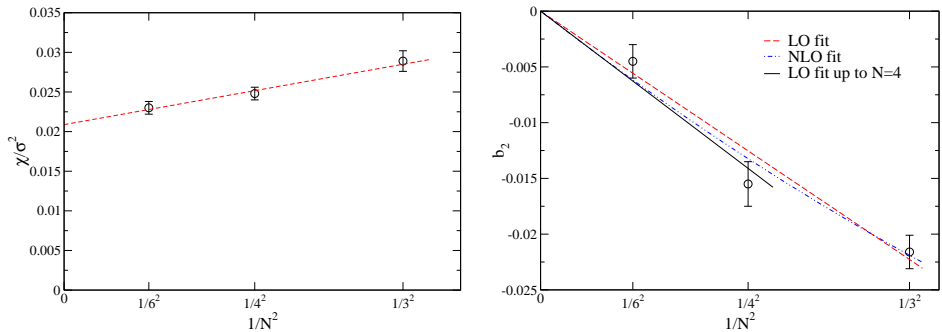

\centering
\includegraphics[width=0.40\textwidth,clip]{./chi_largeN.eps}\quad  
\includegraphics[width=0.395\textwidth,clip]{./b2_largeN.eps}
\caption{(left) Large $N$ limit of the dimensionless ratio $\chi/\sigma^2$;
(right) large $N$ limit of $b_2$.}
\label{fig:largeNres}
\end{figure}

In Fig.~\ref{fig:largeNres} (left) we report the $N$ dependence of the
continuum extrapolated ratio $\chi/\sigma^2$, that nicely follows the
theoretical expectations. By using the scaling form in Eq.~\eqref{largeNsun} we
obtained $(\chi/\sigma^2)|_{\infty}=0.0209(11)$, a result compatible with the
previous determinations \cite{chiLargeN} and only slightly more accurate.
Indeed most of the error comes from the string tension and using a different
scale setting observable would be sufficient to significantly improve the final
error; since our main interest was the dimensionless quantity $b_2$ we did not
pursued this investigation any further.

The large $N$ behaviour of the continuum extrapolated values of $b_2$ is shown
in Fig.~\ref{fig:largeNres} (right). Several large $N$ fits have been tested
(generic power-law, leading order of Eq.~\eqref{largeNsun}, next-to-leading
order of Eq.~\eqref{largeNsun}) and the stability of the fits was tested by
discarding the data corresponding to the case $N=3$. All the fits gave
consistent results (see \cite{Bonati:2016tvi} for more details) and we report as our
final estimate the value $\bar{b}_2=-0.23(3)$.  As previously noted no signal
of a non-vanishing $b_4$ value was observed, and assuming the large $N$ scaling
corresponding to Eq.~\eqref{largeNsun} to hold true for $N=4$ we obtain the
upper bound $|\bar{b}_4|\lesssim 0.1$.

\begin{figure}[t]
\centering
\includegraphics[width=0.40\textwidth,clip]{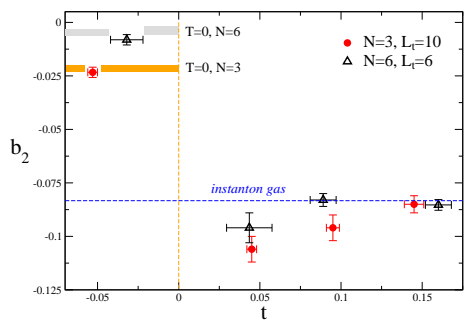}
\caption{Change of large $N$ scaling for $b_2$ across the deconfinement
transition, $t=(T-T_c)/T_c$. Updated version of the figure in
\cite{Bonati:2013tt}.}
\label{fig:finiteT}
\end{figure}

As a final application of the imaginary $\theta$ approach, we greatly improved
the precision of the $b_2$ estimate for $SU(6)$ at finite $T$ presented in
\cite{Bonati:2013tt}: using also data from \cite{Bonati:2015sqt}, the final
figure of \cite{Bonati:2013tt} now becomes Fig.~\ref{fig:finiteT}. With the
new, largely reduced, error bars in the low temperature phase, the change of
the large $N$ scaling across the transition is now even more evident: for
$T<T_c$ the large $N$ limit is governed by the same scaling variable $\theta/N$
as at $T=0$, while for $T>T_c$ no $N$-dependence is observed in $b_2$ and the
functional form $\chi(T)(1-\cos(\theta))$ of the $\theta$ dependence (predicted
by the dilute instanton gas approximation \cite{Gross:1980br}) is quickly
approached.  For $T>T_c$ Eqs.~\eqref{largeNsun} clearly fails: in the
$N\to\infty$ limit $\chi$ vanishes, so there is no reason for $\theta/N$ to be
the relevant scaling variable; the study of $b_{2n}$ shows that the correct
scaling variable is just $\theta$, suggesting that the theory can be described
in terms of effective degrees of freedom carrying an unit of topological
charge.

\section{Conclusions}\label{sec:concl}

In this proceeding we reported on the results obtained in the paper
\cite{Bonati:2016tvi}: a careful investigation of the $\theta$ dependence of
$SU(N)$ gauge theories has been carried out, with the principal aim of studying
the deviations from the leading order quadratic behaviour in $\theta$ of the
ground state energy density. 

In order to carry out such and investigation, the use of simulation performed
at imaginary values of the $\theta$ angle was of paramount importance to reduce
the statistical errors and the systematics related to finite volume effects.
This gave us the possibility of obtaining, for the first time, estimates of
$b_2$ for $N=4, 6$ precise enough to quantitatively verify the large $N$ prediction
in Eq.~\eqref{largeNsun}. Our final estimates for the coefficients governing the 
$\theta$ dependence at $T=0$ in the large $N$ limit are:
\begin{equation}\label{final}
(\chi/\sigma^2)|_{\infty}=0.0209(11),\quad
\bar{b}_2=-0.23(3), \quad
|\bar{b}_4|\lesssim 0.1\ .
\end{equation}

Several recent studies (see e.g. \cite{fracinst}) have given new vigor to the
idea that confinement is related to semiclasical objects carrying fractional
topological charge $1/N$. If the effective interactions between these objects
is small enough, one can proceed as in the dilute instanton approximation and
obtain for the $\theta$ dependence of the ground state energy density the
functional form $E(\theta)-E(0)=\chi(T)(1-\cos(\theta/N))$. This expression
correctly reproduce the large $N$ behaviour of Eq.~\eqref{largeNsun} and
observed in \cite{chiLargeN} and \cite{Bonati:2016tvi}, however its
predictions are not quantitatively accurate: from this functional form, the
value $-1/12\simeq -0.0833$ follows for the $\bar{b}_2$ coefficients, that is not
compatible with the numerical result in Eq.~\eqref{final}. For the case of the
$CP^{N-1}$ models, where similar ideas apply and in which the $\bar{b}_{2n}$s are
analytically known \cite{Bonati:2016tvi}, the situation is even worst: the
$\bar{b}_{2n}$s have all the same sign, while the instanton-like expression
predicts an alternating series. Whether these discrepancies are due to a
different underlying confinement mechanism or to interactions between the
effective degrees of freedom is a question that cannot at present be settled
and surely deserves further studies.

\acknowledgments C.B. and M.D'E. thank Ariel Zhitnitsky for interesting
discussions at the ECT$^*$ workshop ``Gauge topology: from lattice to
colliders''.


\begin{thebibliography}{99}
\bibitem{Vicari:2008jw} 
  E.~Vicari and H.~Panagopoulos,
  %``Theta dependence of SU(N) gauge theories in the presence of a topological term,''
  Phys.\ Rept.\  {\bf 470}, 93 (2009)
  [arXiv:0803.1593 [hep-th]].

\bibitem{DiVecchia:1980ve}
  P.~Di Vecchia and G.~Veneziano,
  %``Chiral Dynamics in the Large n Limit,''
  Nucl.\ Phys.\ B {\bf 171} (1980) 253.

\bibitem{Gross:1980br} 
  D.~J.~Gross, R.~D.~Pisarski and L.~G.~Yaffe,
  %``QCD and Instantons at Finite Temperature,''
  Rev.\ Mod.\ Phys.\  {\bf 53}, 43 (1981).

\bibitem{WittenLargeN} 
  E.~Witten,
  %``Large N Chiral Dynamics,''
  Annals Phys.\  {\bf 128}, 363 (1980);
  E.~Witten,
  %``Theta dependence in the large N limit of four-dimensional gauge theories,''
  Phys.\ Rev.\ Lett.\  {\bf 81}, 2862 (1998)
  [hep-th/9807109].

\bibitem{chiLargeN} 
  B.~Lucini and M.~Teper,
  %``SU(N) gauge theories in four-dimensions: Exploring the approach to N = infinity,''
  JHEP {\bf 0106}, 050 (2001)
  [hep-lat/0103027];
  L.~Del Debbio, H.~Panagopoulos and E.~Vicari,
  %``theta dependence of SU(N) gauge theories,''
  JHEP {\bf 0208}, 044 (2002)
  [hep-th/0204125];
  B.~Lucini, M.~Teper and U.~Wenger,
  %``Topology of SU(N) gauge theories at T =~ 0 and T =~ T(c),''
  Nucl.\ Phys.\ B {\bf 715}, 461 (2005)
  [hep-lat/0401028].

\bibitem{Bonati:2015sqt} 
  C.~Bonati, M.~D'Elia and A.~Scapellato,
  %``$\theta$ dependence in $SU(3)$ Yang-Mills theory from analytic continuation,''
  Phys.\ Rev.\ D {\bf 93}, 025028 (2016)
  [arXiv:1512.01544 [hep-lat]].

\bibitem{Bonati:2016tvi} 
  C.~Bonati, M.~D'Elia, P.~Rossi and E.~Vicari,
  %``$\theta$ dependence of 4D $SU(N)$ gauge theories in the large-$N$ limit,''
  Phys.\ Rev.\ D {\bf 94}, 085017 (2016)
  [arXiv:1607.06360 [hep-lat]].

\bibitem{MBH}
  A.~Milchev, K.~Binder and D.~W.~Heermann
  %``Fluctuations and Lack of Sef-Averaging in the Kinetics of Domain Growth''
  Z. Phys. B -Condensed Matter {\bf 63}, 521 (1986).

\bibitem{Panagopoulos:2011rb} 
  H.~Panagopoulos and E.~Vicari,
  %``The 4D SU(3) gauge theory with an imaginary $\theta$ term,''
  JHEP {\bf 1111}, 119 (2011)
  [arXiv:1109.6815 [hep-lat]].

\bibitem{Campostrini:1988cy} 
  M.~Campostrini, A.~Di Giacomo and H.~Panagopoulos,
  %``The Topological Susceptibility on the Lattice,''
  Phys.\ Lett.\ B {\bf 212}, 206 (1988).

\bibitem{cooling} 
  B.~Berg,
  %``Dislocations and Topological Background in the Lattice O(3) $\sigma$ Model,''
  Phys.\ Lett.\ B {\bf 104}, 475 (1981);
  Y.~Iwasaki and T.~Yoshie,
  %``Instantons and Topological Charge in Lattice Gauge Theory,''
  Phys.\ Lett.\ B {\bf 131}, 159 (1983);
  S.~Itoh, Y.~Iwasaki and T.~Yoshie,
  %``Stability of Instantons on the Lattice and the Renormalized Trajectory,''
  Phys.\ Lett.\ B {\bf 147}, 141 (1984);
  M.~Teper,
  %``Instantons in the Quantized SU(2) Vacuum: A Lattice Monte Carlo Investigation,''
  Phys.\ Lett.\ B {\bf 162}, 357 (1985);
  E.~M.~Ilgenfritz, M.~L.~Laursen, G.~Schierholz, M.~Muller-Preussker and H.~Schiller,
  %``First Evidence for the Existence of Instantons in the Quantized SU(2) Lattice Vacuum,''
  Nucl.\ Phys.\ B {\bf 268}, 693 (1986).

\bibitem{smoothers} 
  C.~Bonati and M.~D'Elia,
  %``Comparison of the gradient flow with cooling in $SU(3)$ pure gauge theory,''
  Phys.\ Rev.\ D {\bf 89}, 105005 (2014)
  [arXiv:1401.2441 [hep-lat]];
  K.~Cichy, A.~Dromard, E.~Garcia-Ramos, K.~Ottnad, C.~Urbach, M.~Wagner, U.~Wenger and F.~Zimmermann,
  %``Comparison of different lattice definitions of the topological charge,''
  PoS LATTICE {\bf 2014}, 075 (2014)
  [arXiv:1411.1205 [hep-lat]];
  Y.~Namekawa,
  %``Comparative study of topological charge,''
  PoS LATTICE {\bf 2014}, 344 (2015)
  [arXiv:1501.06295 [hep-lat]];
  C.~Alexandrou, A.~Athenodorou and K.~Jansen,
  %``Topological charge using cooling and the gradient flow,''
  Phys.\ Rev.\ D {\bf 92}, 125014 (2015)
  [arXiv:1509.04259 [hep-lat]];
  B.~A.~Berg and D.~A.~Clarke,
  %``Deconfinement, gradient and cooling scales for pure SU(2) lattice gauge theory,''
  arXiv:1612.07347 [hep-lat].

\bibitem{Bonati:2013tt} 
  C.~Bonati, M.~D'Elia, H.~Panagopoulos and E.~Vicari,
  %``Change of θ Dependence in 4D SU(N) Gauge Theories Across the Deconfinement Transition,''
  Phys.\ Rev.\ Lett.\  {\bf 110}, no. 25, 252003 (2013)
  [arXiv:1301.7640 [hep-lat]].

\bibitem{fracinst} 
  M.~Unsal and L.~G.~Yaffe,
  %``Center-stabilized Yang-Mills theory: Confinement and large N volume independence,''
  Phys.\ Rev.\ D {\bf 78}, 065035 (2008)
  [arXiv:0803.0344 [hep-th]];
  A.~Parnachev and A.~R.~Zhitnitsky,
  %``Phase Transitions, theta Behavior and Instantons in QCD and its Holographic Model,''
  Phys.\ Rev.\ D {\bf 78}, 125002 (2008)
  [arXiv:0806.1736 [hep-ph]].


\end{thebibliography}
\end{document}